\documentclass[aps,prd,epsf,showpacs,amsmath,amssymb,graphics,
twocolumn,10pt]{revtex4}
\usepackage{graphicx}
\usepackage{dcolumn}
\usepackage{bm}

\newcommand{\be}{\begin{equation}}
\newcommand{\ee}{\end{equation}}
\newcommand{\ba}{\begin{eqnarray}}
\newcommand{\ea}{\end{eqnarray}}
\newcommand{\ban}{\begin{eqnarray*}}
\newcommand{\ean}{\end{eqnarray*}}

\newcommand{\eq}[1]{(\ref{#1})}

\begin{document}
\title{Circular geodesics and accretion disks in Janis-Newman-Winicour and Gamma metric}

\author{Anirban N. Chowdhury} \email{a.chowdhury@iiserpune.ac.in }
\affiliation{Tata Institute of Fundamental Research, Homi Bhabha
Road, Colaba, Mumbai 400005, India}
\author{Mandar Patil} \email{mandarp@tifr.res.in} \affiliation{Tata
Institute of Fundamental Research, Homi Bhabha Road, Colaba, Mumbai 400005, India}
\author{Daniele Malafarina} \email{daniele.malafarina@polimi.it}
\affiliation{Tata Institute of Fundamental Research, Homi Bhabha Road,
Colaba, Mumbai 400005, India}
\author{Pankaj S. Joshi} \email{psj@tifr.res.in} \affiliation{Tata
Institute of Fundamental Research, Homi Bhabha Road, Colaba,
Mumbai 400005, India}


\begin{abstract}
We study here circular timelike geodesics in the Janis-Newman-Winicour
and Gamma metric spacetimes which contain a strong curvature naked singularity
and reduce to the Schwarzschild metric for a specific value of one
of the parameters.
We show that for both the metrics the range of allowed parameters
can be divided into three regimes where structure of the circular
geodesics is qualitatively different.
It follows that the properties of the accretion disks around
such naked singularities can be significantly different from
those of disks around black holes. This adds to previous studies
showing that if naked singularities exist in nature, their
observational signature would be significantly different from
that of the black hole.
\end{abstract}

\pacs{04.20.Dw,04.20.Jb,04.70 Bw}
\keywords{Gravitational collapse, black holes, naked singularity}

\maketitle

\section{Introduction}

One of the most important unresolved problems in general
relativity, and an issue that has far reaching observational
implications, is that of the final fate of complete gravitational
collapse of a massive body such as a star.

The Israel-Carter conjecture asserts that a Kerr black hole, which is
described by two parameters, namely mass and angular momentum,
would be formed as the endstate of any generic complete gravitational
collapse and thus all other information regarding the nature of the
matter fields, symmetries, initial conditions will be radiated
away in the process of collapse. The above statement remains at
the stage of conjecture as of now because proving it analytically or
numerically is turning out to be a very difficult task.
We also do not understand very well how matter would behave at
energies beyond nuclear energy scales all the way up to Planck scale
as we have no theoretical understanding of evolution of the equation
of state of matter at such energies. Thus the theoretical
studies do not provide a definite answer to the question
whether or not black holes are the only possible outcomes of
the gravitational collapse.

In fact in the last few years new theoretical models and
observations seem to suggest that the true nature of the process
could be more complicated. Investigations of gravitational collapse of
various matter configurations in the context of the general theory
of relativity over last couple of decades has revealed that
the endstates of complete gravitational collapse could either
be black holes or naked singularities
\cite{PSJ}.

Therefore it is natural to ask the question of what
happens during a realistic collapse of a massive body and what
will be the endstate of such a collapse. The possibilities can
be divided into three main categories:
\begin{itemize}
  \item[-] The body radiates away all higher multipole moments and all
asymmetry, thus forming a Kerr black hole.
  \item[-] Collapse will halt before all matter is squeezed
into a spacetime singularity, thus creating a finite sized final
object that is different from a black hole.
  \item[-] Matter will fall into the final spacetime singularity
preserving its symmetry structure, thus creating a final configuration
in the form of vacuum spacetime with a singularity
(that can be covered or naked).
\end{itemize}

The analytical solutions describing dynamical collapse
of a rotating body away from spherical symmetry are few and not well
understood at present. Therefore a possible mechanism to account
for the first hypothesis is still missing. For this reason it seems
natural to turn attention to the other two options.
Ideas following the second hypothesis have been proposed
\cite{Herrera},
but the nature of these objects and the physical processes
that could lead to the formation of the same are not well
understood. Furthermore recently some new scenarios have been
proposed that include the possibility that gravitational collapse
can asymptotically halt, leading to the formation of static
configuration of matter that may or may not contain a naked
singularity and which could be either finite or infinite
in extent
\cite{JNM}.

On the other hand, if black holes and naked singularities
which are hypothetical astrophysical objects occur in nature,
a question would be how they would be observationally
different. Therefore some researchers have started looking
into the observational features that would distinguish these entities.
The black holes and naked singularities could have rather
different properties and this could also possibly shed a light
on the nature of the existing sources, like the supermassive dark
objects that dwell at the center of galaxies.
Recent studies of gravitational lensing
(see for example \cite{lensing})
and accretion disks
(see \cite{JNM,accretion,accretion2} and \cite{Harko})
have brought out interesting characteristic differences
for these objects which could possibly distinguish them
from each other. Recently it
was also suggested by some of us that ultrahigh energy collisions
and fluxes of the escaping collision products could be used
for this purpose in certain situations
\cite{Patil}.

In the present paper we investigate the properties of
circular geodesics and accretion disks in two spacetimes
that arise naturally while generalizing the Schwarzschild solution.
One is the Janis-Newman-Winicour (JNW) solution and the
other one is the so called Gamma metric ($\gamma$-metric
henceforth).

The Schwarzschild metric is the unique vacuum
solution to Einstein equations under spherical symmetry
which is asymptotically flat. Different metrics can be
obtained from the Schwarzschild solution by relaxing one or more
of the assumptions above. Firstly, relaxing the spherical symmetry,
we can consider either a static deformation keeping the axial symmetry intact,
or we could invoke a spin. One of the solutions that can
be obtained in the former case is the $\gamma$-metric
\cite{gamma1},
whereas the spacetime in the latter case is described
by the Kerr metric.
On the other hand the assumption of the empty spacetime
can be relaxed by invoking the presence of either an
electromagnetic field or a scalar field. In the former case
one obtains the Reissner-Nordstr\"{o}m solution, whereas
in the latter case, by invoking a massless scalar field,
the JNW solution is obtained
\cite{JNW1}.

The uniqueness theorems give a privileged status to the Kerr and
Schwarzschild spacetimes. Nevertheless since we do not know if in
a realistic physical scenario those will necessary be the resulting
metrics it is of interest to study spacetimes such as JNW and the
$\gamma$-metric that depart from the black hole models and see 
if they can be observationally different from their black hole 
counterparts.
The main point that makes the $\gamma$-metric and the JNW-metric 
more appealing is that they are continuously connected to the Schwarzschild
solution via the value of one parameter. By reducing this parameter to
a specific value we recover exactly the Schwarzschild solution.
The interesting point is that the Schwarzschild metric has a black
hole whereas these two generalizations contain naked singularities
and therefore we can directly compare the results obtained for models of
accretion disks in these metrics with the Schwarzschild case.
Furthermore the parameters appearing in the JNW and $\gamma$-metric have a clear
and direct physical interpretation, as one represents the addition of a scalar
field while the other represents an axial deformation.

The first step towards the study of accretion disks is
an understanding of the circular geodesics in the
equatorial plane of the spacetime. The circular geodesics in
Kerr and Reissner-Nordstr\"{o}m spacetimes containing black hole
as well as the naked singularities depending on the values
of the spin and charge parameters respectively was recently
carried out in \cite{accretion2}. In this paper we investigate
the structure of circular geodesics for the JNW and
$\gamma$ spacetimes.

We find that the properties of accretion disks can be
divided in three regimes according to the values assumed
by a parameter in the solutions. In one regime, the stable
circular geodesics extend all the way up to singularity,
in the second the stable circular orbits can exist in two disjoint
disks separated by a gap, one disk extending from singularity to
a finite radius and another extending from a larger radius to
infinity. Finally, in the third case a photon sphere surrounds
the singularity, and stable circular geodesics can exist
from a particular radius above photon sphere to infinity.

\section{Circular Equatorial Geodesics in the JNW Metric}
In this section we investigate the structure of the circular geodesics
in the JNW spacetime. As mentioned earlier the JNW metric is obtained
as an extension of the Schwarzschild spacetime when a massless
scalar field is present. In a naive sense, the black hole event horizon,
{\it i.e.} a coordinate singularity in the Schwarzschild spacetime
transforms to a real physical singularity which is naked. The JNW metric
depends upon two parameters, one related to the mass and an additional
parameter which can be interpreted as the strength of the scalar
field or `scalar charge'.

The JNW metric can be written in the following way
\begin{equation}
ds^2=-Adt^2+A^{-1}dr^2+Bd\Omega^2 \; ,
\label{JNWM1}
\end{equation}
where $d\Omega^2=d\vartheta^2+\sin^2\theta d\phi^2$ is the
line element of a unit two-sphere and the functions $A(r)$ and $B(r)$
are given by the following expressions
\begin{eqnarray}
  A(r) &=& \left[\frac{2r-r_0 (\mu-1)}{2r+r_0 (\mu+1)}\right]^\frac{1}{\mu} \; , \\
  B(r) &=& \frac{1}{4}\frac{[2r+r_0 (\mu+1)]^{\frac{1}{\mu}+1}}{[2r-r_0 (\mu-1)]^{\frac{1}{\mu}-1}} \; .
  \label{JNWM2}
\end{eqnarray}
The two parameters mentioned above are $r_0$ (that is related to
the mass) and the scalar charge $\mu$. The range of $\mu$ is given
by $\mu \in (1,\infty)$ and he scalar field is given by the
following expression,
\begin{align}
&\varphi=\frac{a}{\mu}\ln\left|\frac{2r-r_0 (\mu-1)}{2r+r_0 (\mu+1)}\right| \; ,
\end{align}
where $\mu$ and $a$ are linked by the relation
$\mu = \left(1+8\pi\frac{4a^2}{r_0^2} \right)$.
The JNW spacetime contains a strong curvature singularity which is globally naked at
\begin{equation}
r_{sing}=\frac{1}{2}r_0 (\mu-1) \; ,
\end{equation}
 for $\mu\neq1$.

The Schwarzschild solution in the usual form can be recovered from the JNW solution when $\mu=1$ after a simple coordinate transformation of the radial coordinate given by $\bar{r}=r+r_0$, with $r_0=2M$
which is the usual Schwarzschild radius of the event horizon (note that the radial coordinate $r$ of the JNW metric when $\mu=1$ does not correspond to the usual Schwarzschild coordinate $\bar{r}$), and where $M$ is related to the total mass of the central body as
measured by observers at infinity.
Therefore the case where $\mu$ is close to $1$ corresponds to the small deviations from the Schwarzschild metric due to the presence of a small massless scalar field, whereas large values of $\mu$ correspond to the large deviations from the Schwarzschild geometry
\cite{JNW2}.

\subsection{Geodesics in JNW spacetime: Basic equations}
We now investigate the structure of timelike equatorial geodesics in the JNW spacetime.
The JNW metric is spherically symmetric and static and therefore admits the two Killing vectors $\xi_t=\partial_{t}$ and $\xi_\phi=\partial_{\phi}$.
As a consequence we have the following conserved quantities along geodesic motion.
\begin{align} \label{LEJNW1a}
&E=-g_{\alpha\beta}\xi_t^\alpha u^\beta=A(r)U^t \; , \\ \label{LEJNW1b}
&L=g_{\alpha\beta}\xi_\phi^\alpha u^\beta=B(r)U^{\phi} \; ,
\end{align}
where $U^{\alpha}=(U^{t},U^{r},U^{\theta},U^{\phi})$ is the four-velocity of the test particle. The constants of motion $E$ and $L$ can be interpreted as the conserved energy and orbital angular momentum per unit mass of the particle respectively.

Spherical symmetry implies that the motion of the particle will be restricted to a plane which can be chosen by gauge freedom to be the equatorial plane. Thus we set $\theta=\frac{\pi}{2}$ and $U^{\theta}=0$.

From equations \eq{LEJNW1a} and \eq{LEJNW1b} using the normalization condition for the four-velocity $U^{\alpha}U_{\alpha}=-1$, we obtain the equation for the motion of a particle in the radial direction.
\begin{equation}
\dot{r}^2+V^2=E^2 \; .
\label{JNWrd1}
\end{equation}
Here $V$ can be thought of as an effective potential for the test particle and it is given by a following expression,
\begin{equation}
V^2=A(r)\left(1+\frac{L^2}{B(r)}\right) \; .
\label{JNWVef1}
\end{equation}

\subsection{Circular geodesics}
From the analysis of the above effective potential we can obtain the structure of circular geodesics.
A particle orbiting along a circular geodesics must obey the conditions
\begin{equation}
U^{r}=\dot{U}^{r}=0 \; ,
\label{JNWcon1}
\end{equation}
as $r$ must stay constant.
Here the dot denotes a derivative with respect to
the affine parameter for geodesic {\it i.e.} it is
the proper time.

From equations \eq{JNWrd1} and \eq{JNWcon1} it follows that for circular geodesics
\begin{equation}
\label{JNWcond1}
V=E \; , \; \text{and} \; \frac{\partial V}{\partial r}=0 \; .
\end{equation}
Further one needs to ensure that the circular orbit is stable against the small perturbations in the radial direction and this implies that for a stable orbit the effective potential must admit a minimum:
\begin{equation}
\frac{\partial^2 V}{\partial r^2}>0 \; .
\label{JNWstab1}
\end{equation}
When the effective potential admits a maximum i.e. when $\frac{\partial^2 V}{\partial r^2}<0$ the orbit is unstable against small perturbations.
The condition for the marginally stable orbit is
$\frac{\partial^2 V}{\partial r^2}=0$,
corresponding to the inflection point for the effective potential.

Imposing the conditions \eq{JNWcond1} on \eq{JNWVef1} we get the expression for the conserved energy and momentum of the particle orbiting along the geodesic of radius $r$ as
\begin{align}
\label{JNWLE1}
&L^2=\frac{r_0 }{4(2r-r_0)}\frac{[2r+r_0 (\mu+1)]^{\frac{1}{\mu}+1}}{[2r-r_0 (\mu-1)]^{\frac{1}{\mu}-1}} \; ,\\
\label{JNWLE2}
&E^2=\frac{2r}{2r-r_0}\left[\frac{2r-r_0 (\mu-1)}{2r+r_0 (\mu+1)}\right]^\frac{1}{\mu} \; ,
\end{align}
from which it follows that we must have
\begin{equation}
r>\frac{r_0}{2}=M \; ,
\end{equation}
for the energy and angular momentum to be real. Thus the circular geodesics can exists only for those values of $r$ for which this condition holds.
The energy and angular momentum diverge at the radius
\begin{equation}
r_{ph}=\frac{r_0}{2} \; ,
\end{equation}
which represents the radius at which photons move on circular trajectories and it is known as photon sphere. The photon sphere plays a crucial role in the study of gravitational lensing since its presence delimits a boundary below which lensing effects cannot be observed, therefore implying that two spacetimes that possess a photon sphere at the same radius cannot be distinguished by the observation of lensing effects even though their compact sources might be different.

Imposing the condition for the stability of the circular geodesics against small perturbations along the radial direction in equation \eq{JNWVef1} we get
\begin{align}
&4r^2-8rr_0+r_0^2(\mu^2-1)>0 \; ,
\end{align}
which implies that
the radius of the stable orbit must satisfy the following condition
\begin{eqnarray}
r>r_+ \; \text{or} \;  r<r_- \; ,
\end{eqnarray}
where
\begin{align}\label{rpm}
&r_\pm=r_0(1\pm \frac{1}{2}\sqrt{5-\mu^2}) \; .
\end{align}
The marginally stable orbits are located at these two values of the radii.

\begin{figure}
\includegraphics[width=0.5\textwidth]{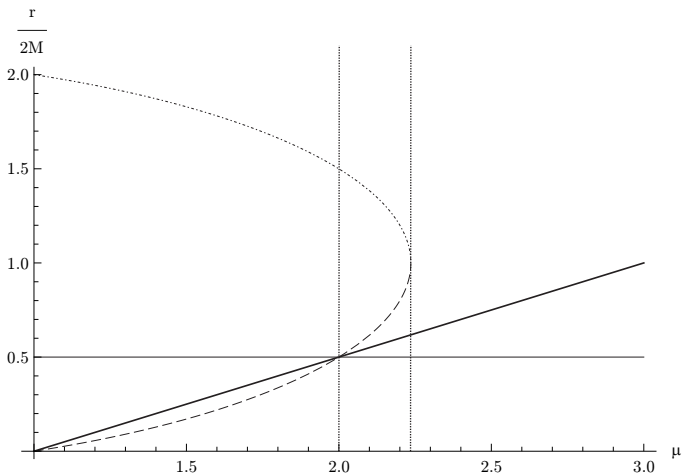}
\caption{\label{Fig1}
The value of the radial coordinate (in units of $\frac{1}{2M}$) for the singularity (thick line), photon sphere (thin line), $r_+$ (dotted and dashed line) and $r_-$ (dashed line) as functions of the parameter $\mu$. It is clear that the structure of the circular geodesics will be qualitatively different in the three different ranges of the parameter, namely (1,2),(2,$\sqrt{5}$) and ($\sqrt{5}$,$\infty$).}
\end{figure}

In Fig. \ref{Fig1} are shown the values of the radial coordinates for the singularity, photon sphere, $r_+$ and $r_-$ depending upon the parameter $\mu$. Three distinct ranges of the parameters emerge where the structure of the circular geodesics will be qualitatively different, namely $\mu \in (1,2)$, $\mu \in (2,\sqrt{5})$, $\mu \in (\sqrt{5},\infty)$. We discuss these cases separately.

\begin{figure}
\includegraphics[width=0.5\textwidth]{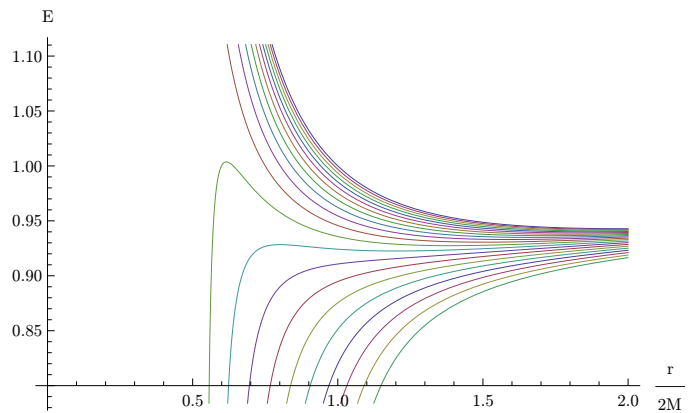}
\caption{\label{Fig2}
The conserved energy per unit mass $E$ is plotted against the radius of the circular orbit $r$ (in units of $\frac{1}{2M}$) for different values of $\mu$. The curves located higher in the diagram correspond to lower values of $\mu$ and vice versa.
For values of $\mu \in (1,2)$, $E$ decreases with the decreasing $r$, attains a minimum and again increases to become infinite before the singularity is reached. For values of $\mu \in (2,\sqrt{5})$, $E$ initially decreases with decreasing $r$, attains a minimum followed by a maximum and decreases to a zero value at the singularity. Whereas for large values of $\mu \in (\sqrt{5},\infty)$, $E$ is monotonically decreasing with decreasing $r$ and takes a zero value at the singularity.}
\end{figure}

\begin{figure}
\includegraphics[width=0.5\textwidth]{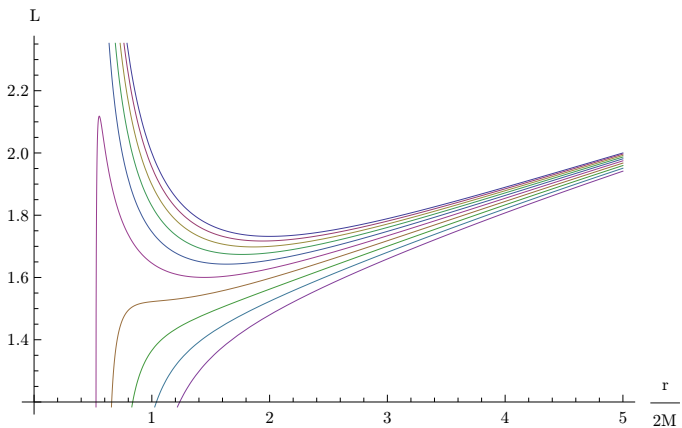}
\caption{\label{Fig3}
The conserved angular momentum per unit mass $L$ is plotted against the radius of the circular orbit $r$ (in units of $\frac{1}{2M}$) for different values of $\mu$. The curves located higher in the diagram correspond to lower values of $\mu$ and vice versa.
For values of $\mu \in (1,2)$, $L$ has a minimum at a finite $r$ and increases diverging as the singularity is approached and as $r$ goes to infinity. For values of $\mu \in (2,\sqrt{5})$ $L$ vanishes at the singularity and has a local maximum and a local minimum before increasing indefinitely as $r$ goes to infinity. Whereas for large values of $\mu \in (\sqrt{5},\infty)$, $L$ increases monotonically from zero at the singularity to infinity as $r$ goes to infinity.}
\end{figure}

The conserved energy and angular momentum for test particles as a function of the radius of the circular orbit are plotted in Fig. \ref{Fig2}and Fig.\ref{Fig3} respectively, for different values of the parameter $\mu$. The qualitative behaviour of these functions is also manifestly different in the three regimes of the parameter mentioned above.

\subsubsection{Case $\mu \in (1,2)$}
For this range of values of parameters we have a following relation, as seen from Fig. \ref{Fig1},
\begin{equation}
r_+ >r_{ph}>r_{sing}>r_{-} \; .
\end{equation}
This implies that $r_-$ does not exist and a photon sphere covers the singularity. The stable circular orbits can exist in the region given by,
\begin{equation}
    r \in (r_{+},\infty) \; ,
\end{equation}
which is above the photon sphere. Unstable circular orbits can exist between the photon sphere and $r_{+}$ and no circular orbits can exists below the photon sphere up to the singularity.

\begin{figure}
\includegraphics[width=0.5\textwidth]{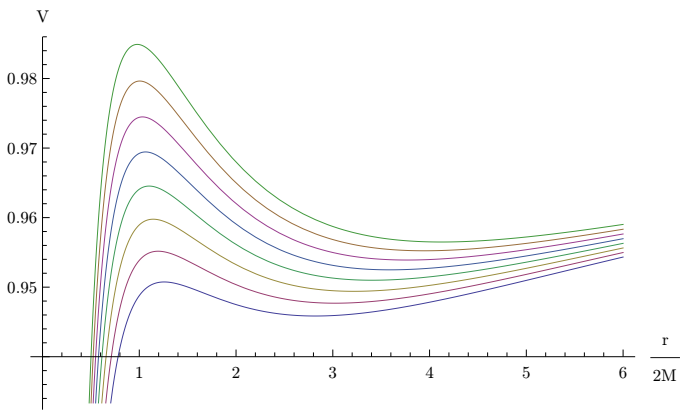}
\caption{\label{Fig4}
The effective potential for radial motion $V$ is plotted as a function of radial coordinate $\frac{r}{2M}$ for $\mu=1.5 \in (1,2)$ for different values of $L$. It can be seen that the $V$ admits one minimum and one maximum, which corresponds to stable and unstable circular orbits respectively.
}
\end{figure}

In Fig.\ref{Fig4} the general behaviour of the effective potential is shown as a function of the radius for a fixed value of the parameter $\mu$ in the range considered and for different values of the angular momentum $L$. The effective potential admits
one maximum and one minimum which correspond to the unstable and stable circular orbits, respectively. The unstable circular orbits lie in the region between the photon sphere and $r_{+}$.

Note that both $E(r)$ and $L(r)$ become infinite at a certain finite radius $r$ in this case (as can be seen from Figs. \ref{Fig2} and \ref{Fig3}).
As we approach the singularity from infinity, the stable circular geodesics can exist up to the minimum of energy and angular momentum. The unstable orbits exist in the region between the minimum and the radial value where $E(r)$ and $L(r)$ blow up, which is the photon sphere.
No circular orbits can exist below the photon sphere a result that is consistent with the earlier discussion.

\subsubsection{Case $\mu \in (2,\sqrt{5})$}
In this range of parameters we get the following relation
\begin{equation}
r_+ >r_{-}>r_{sing}>r_{ph} \; .
\end{equation}
This implies that the photon sphere does not exist and there are two distinct regions where circular geodesics can exist namely,
\begin{equation}
r \in (r_{+},\infty) \; \text{and} \;
r \in (r_{sing},r_{-}) \; .
\end{equation}
A new interesting feature arises in this case since
the second region extends all the way up to singularity.
Particles reaching the innermost boundary of the exterior accretion disk would plunge in the region of unstable orbits
to reach $r_-$ where they would shock and circularize again, inspiralling onto the singularity.

\begin{figure}
\includegraphics[width=0.5\textwidth]{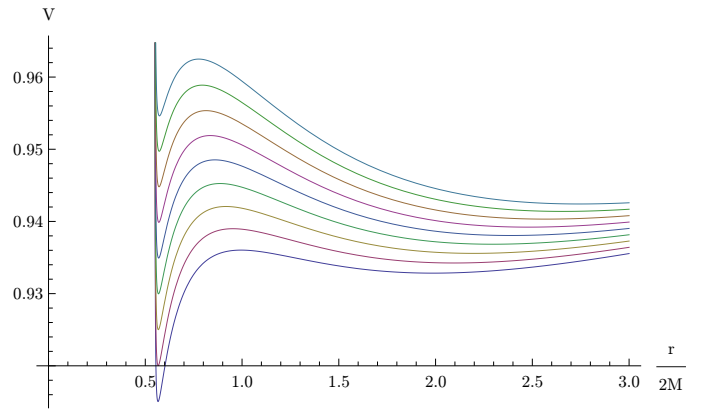}
\caption{\label{Fig5}
The effective potential for radial motion $V$ is plotted as a function of radial coordinate $\frac{r}{2M}$ for $\mu=2.1 \in (2,\sqrt{5})$ for different values of $L$. It can be seen that $V$ admits two minima and one maximum in between, which indicate the presence of two regions with stable circular orbits separated by one region of unstable circular orbits.}
\end{figure}

In Fig.\ref{Fig5} the general behaviour of the effective potential for this range of the parameter is shown for different values of $L$. The effective potential admits two minima which correspond to the two stable circular orbits in the two regions we discussed above. There is also one local maximum between the two minima, which lies in the unstable region between the two stable ones.

The qualitative behaviour of the functions $E(r)$ and $L(r)$ in this range of parameter shows that both quantities do not blow up, instead as $r$ decreases they reach a maximum and then decrease to zero as singularity is approached (see Figs. \ref{Fig2} and \ref{Fig3}). The existence of two regions where stable circular geodesics can exist can be derived also from the analysis of $E(r)$ and $L(r)$ as the two regions correspond to the portions where the two functions decrease with the decreasing $r$, namely from infinity to the first minimum and from the maximum to the singularity.
Since the two functions do not blow up at any allowed radius, no photon sphere is present in the spacetime.

\subsubsection{Case $\mu \in (\sqrt{5},\infty)$}
Within this range of values for the parameter $r_+$ and $r_-$ do not exist as the solutions of equation \eqref{rpm} are not real. For this range of parameters we have
\begin{equation}
r_{sing}>r_{ph} \; ,
\end{equation}
from which we see that the photon sphere is not present as well.
This implies that the stable circular orbits can exist for
\begin{eqnarray}
r \in (r_{sing},\infty) \; ,
\end{eqnarray}
{\it i.e.}from the singularity to infinity everywhere on the equatorial plane.

\begin{figure}
\includegraphics[width=0.5\textwidth]{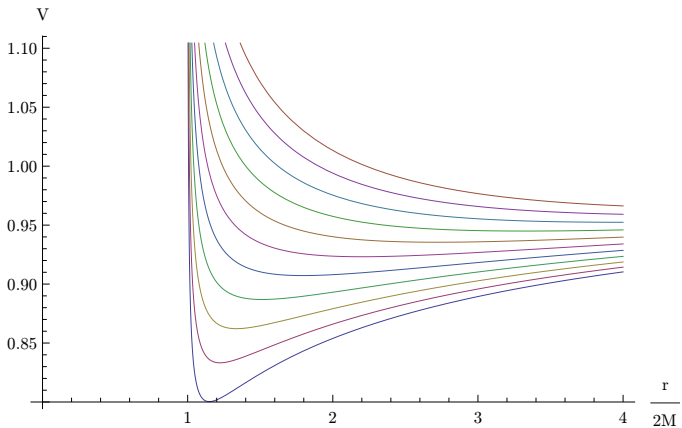}
\caption{\label{Fig6}
The effective potential for radial motion $V$ is plotted against $r$ (in units of $\frac{1}{2M}$) for $\mu=3 \in (\sqrt{5},\infty)$ for different values of $L$. The effective potential admits only one minimum. Unstable orbits do not exists since $V$ does not admit maximum.}
\end{figure}

In Fig.6 the general behaviour of the effective potential for this range of parameter values for various angular momenta is shown. The effective potential admits only one minimum which indicates the presence of stable circular orbit. Since the potential does not admit maximum, unstable circular orbits do not exists.
The circular orbits exist all the way up to the singularity.

In this case the qualitative behaviour of the functions $E(r)$ and $L(r)$ is decreasing from infinity until they become zero at the singularity (see Figs. \ref{Fig2} and \ref{Fig3}). This implies that the stable circular orbits would exist all the way up to singularity, which is consistent with the above discussion.

\section{Circular Geodesics in the Gamma metric}
In this section we study the structure of the circular geodesics in the $\gamma$ spacetime. The $\gamma$-metric is obtained by extending the Schwarzschild spacetime when spherical symmetry is deformed to develop a prolate or oblate spheroid.

The $\gamma$-metric is a two parameter family of spacetimes belonging to the Weyl class of static, axially symmetric, vacuum solutions of Einstein equations that are asymptotically flat. In principle all the metrics belonging to the Weyl class are known since there is a one to one correspondence between such spacetimes and solutions of the Laplace equation in flat two-dimensional space. The interesting feature of the $\gamma$-metric is that it is continuously linked to the Schwarzschild metric via one of the parameters.
The two parameters can be defined as a mass parameter $M$ and the parameter $\gamma$, that quantifies the deformation from spherical symmetry. We investigate the circular geodesics in $\gamma$-spacetime as a function of the deformation parameter.

The $\gamma$-metric is given by
\begin{align}
&ds^2=-Fdt^2+F^{-1}[Gdr^2+Hd\theta^2+(r^2-2mr)\sin^2 \theta d\phi^2] \; ,
\label{gamma1}
\end{align}
where the functions $F(r)$, $G(r,\theta)$ and $H(r,\theta)$ are given by the following expressions
\begin{align}
\label{gamma2}
&F=\left(1-\frac{2M}{r}\right)^\gamma \; , \\
&G=\left(\frac{r^2-2Mr}{r^2-2Mr+M^2\sin^2 \theta}\right)^{\gamma^2-1} \; ,\\
&H=\frac{(r^2-2Mr)^{\gamma^2}}{(r^2-2Mr-M^2\sin^2 \theta)^{\gamma^2-1}} \; .
\end{align}
The total ADM mass as measured by an observer at infinity associated with the $\gamma$-spacetime is given by $M_{tot}=\gamma M$ and the range of the deformation parameter is over all the positive real numbers with the Schwarzschild spacetime recovered when $\gamma=1$, whereas $\gamma>1$ ($\gamma<1$) corresponds to an oblate(prolate) geometry respectively. In the limit $\gamma=0$ a flat spacetime is obtained.

There is a strong curvature singularity in the $\gamma$ spacetime whenever $\gamma\neq 1$ and it is located at the radial value
\begin{equation}
r_{sing}=2M \; .
\end{equation}
The singularity is visible to all observers if $\gamma<1$ while for $\gamma>1$ it has a directional behaviour being visible only to observers not on the polar plane. So if we confine ourselves to the equatorial plane, the singularity is a naked singularity
\cite{gamma2}.

\subsection{Geodesics in $\gamma$-spacetime: basic equations}
The complete structure of geodesic motion in the $\gamma$-metric was studied by Herrera et al.
\cite{Herrera2}.
The $\gamma$-metric is axially symmetric and static. Thus as in the case of JNW metric the existence of the Killing vectors $\xi_{t}=\partial_t$ and $\xi_{\phi}=\partial_{\phi}$ implies that the following two quantities are constants of motion
\begin{align}
\label{elgamma1}
&E=-g_{\alpha\beta}\xi_t^\alpha U^\beta=FU^t \; ,\\
&L=g_{\alpha\beta}\xi_\phi^\alpha U^\beta=\frac{r^2-2Mr}{F}U^{\phi} \; ,
\label{elgamma2}
\end{align}
where $U^{\alpha}=\left(U^{t},U^{r},U^{\theta},U^{\phi}\right)$. We assume that the geodesics are confined to the equatorial plane and thus impose $\theta=\frac{\pi}{2}$.

Using the condition for the normalization of velocity $U^{\alpha}U_{\alpha}=-1$ and equations \eq{elgamma1} and \eqref{elgamma2} the equation for the radial motion can be written as
\begin{align}
&G \dot{r}^2+V^2=E^2 \; ,
\end{align}
where $V$ can be thought of as an effective potential and it is given by the following expression
\begin{align}
&V^2\equiv F+\frac{L^2 F^2}{r^2-2Mr} \; .
\label{ve1}
\end{align}

\subsection{Circular geodesics}
As described in the previous section the conditions for the circular geodesic are given by
\begin{equation}
U^{r}=\dot{U}^r=0 \; ,
\end{equation}
where the dotted quantity means derivative with respect to the affine parameter, {\it i.e.} the proper time.
These again are equivalent to
\begin{equation}
\label{con1}
V=E \; , \; \text{and} \; \frac{\partial V}{\partial r}=0 \; ,
\end{equation}
meaning that the effective potential must be equal to the energy and it must admit an extremum.

For the circular orbits to be stable against radial perturbations the effective potential must admit a minimum, thus satisfying the condition
\begin{equation}
\label{con2}
\frac{\partial^2 V}{\partial r^2}>0 \; .
\end{equation}
For the unstable circular geodesics we have $\frac{\partial^2 V}{\partial r^2}<0$, whereas for the marginally stable case we get $\frac{\partial^2 V}{\partial r^2}=0$.

Now imposing condition \eq{con1} on the effective potential \eq{ve1} we obtain the following expressions for angular momentum and energy
\begin{align}
\label{con3}
&L^2=\left(1-\frac{2M}{r}\right)^{1-\gamma}\frac{r^2M\gamma}{r-M(1+2\gamma)} \; , \\
&E^2=\left(1-\frac{2M}{r}\right)^{\gamma}\frac{r-M(1+\gamma)}{r-M(1+2\gamma)} \; .
\end{align}
Circular orbits exist if the condition $r\geq M(1+2\gamma)$ is satisfied which ensures that $E$ and $L$ are real.

Conserved energy and angular momentum blow up at the radius of the photon sphere which is given by the equation
\begin{equation}
r_{ph}=M(1+2\gamma) \; .
\end{equation}

The condition for the stability of the circular geodesics from equations is then
\begin{equation}
r>r_+ \; \text{or} \; r<r_- \; ,
\end{equation}
where
\begin{equation}
r_{\pm}=M\left(1+3\gamma \pm \sqrt{5\gamma^2-1}\right) \; ,
\end{equation}
are the radii of marginally stable orbits.

\begin{figure}
\includegraphics[width=0.5\textwidth]{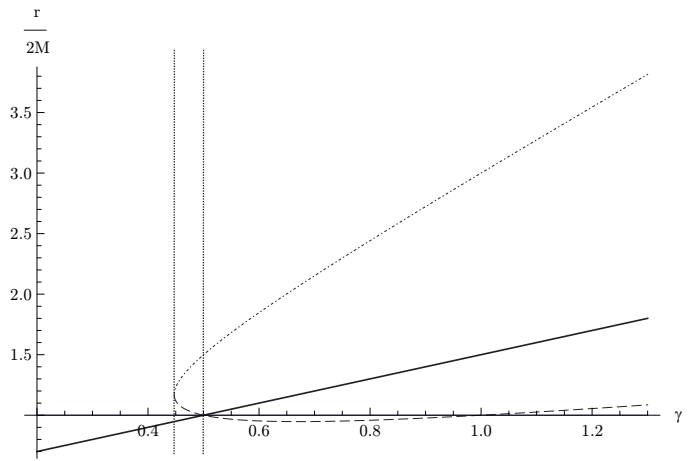}
\caption{\label{Fig7}
The radial curves (in units of $\frac{1}{2M}$) for the singularity (x-axis), photon sphere (thick line), $r_+$ (dotted and dashed line) and $r_-$ (dashed line) respectively as functions of the parameter $\gamma$. It is obvious from the graphs that the structure of the circular geodesics will be qualitatively different in the three different regimes of the parameter, namely (0,$\frac{1}{\sqrt{5}}$), ($\frac{1}{\sqrt{5}}$, $\frac{1}{2}$) and ($\frac{1}{2}$,$\infty$).}
\end{figure}

In Fig. \ref{Fig7} are shown the values of radial coordinate for the singularity, photon sphere, $r_+$ and $r_-$ as the parameter $\gamma$ is varied. Three distinct ranges of the parameters emerge where the structure of the circular geodesics will be qualitatively different, namely $\gamma \in (0,\frac{1}{\sqrt{5}})$, $\gamma \in (\frac{1}{\sqrt{5}},\frac{1}{2})$ and $\gamma \in (\frac{1}{2},\infty)$. We discuss these cases separately.

\begin{figure}
\includegraphics[width=0.5\textwidth]{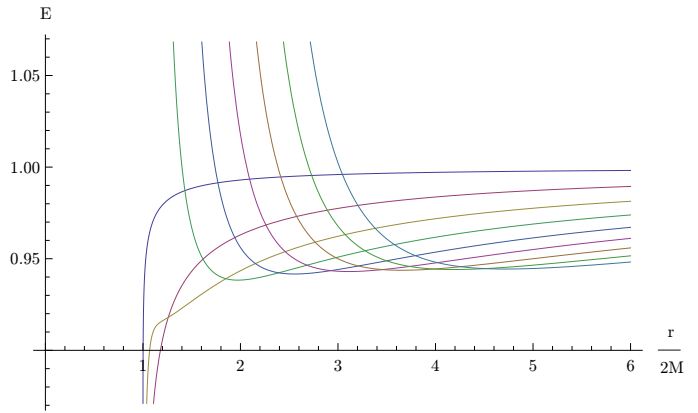}
\caption{\label{Fig8}
The conserved energy per unit mass $E$ is plotted against the radius of the circular orbit $r$ (in units of $\frac{1}{2M}$)for different values of $\gamma$. The curves located higher in the diagram correspond to higher values of $\gamma$.
For values of $\gamma \in (\frac{1}{2},\infty)$, $E$ diverges at the singularity, decreases to a minimum at a finite radius and then increases to diverge again as $r$ goes to infinity. For values of $\gamma \in (\frac{1}{\sqrt{5}},\frac{1}{2})$, $E$  goes to zero at the singularity and it has a maximum and a minimum. Whereas for values of $\gamma \in (0,\frac{1}{\sqrt{5}})$, $E$ is monotonically increasing from zero at the singularity to infinite as $r$ goes to infinity.}
\end{figure}

\begin{figure}
\includegraphics[width=0.5\textwidth]{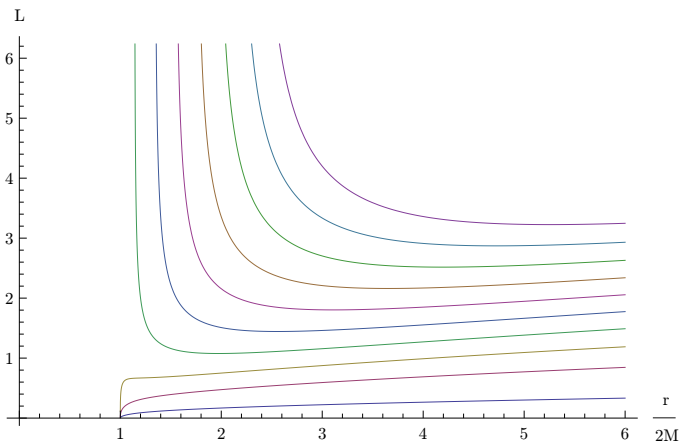}
\caption{\label{Fig9}
The conserved angular momentum per unit mass $L$ is plotted against
the radius of the circular orbit $r$ (in units of $\frac{1}{2M}$) for
different values of $\gamma$.
The curves located higher in the diagram correspond to higher values of $\gamma$.
For values of $\gamma \in (\frac{1}{2},\infty)$ $L$ diverges at the
singularity and as $r$ goes to infinity, thus having a minimum at a
finite radius. For values of $\gamma \in
(\frac{1}{\sqrt{5}},\frac{1}{2})$ $L$ has both a minimum
and a maximum, goes to zero as $r$ approaches the singularity and
diverges as $r$ goes to infinity.
Whereas for values of $\gamma \in (0,\frac{1}{\sqrt{5}})$ $L$ is
monotonically increasing from zero at the singularity to
infinity as $r$ grows indefinitely.}
\end{figure}

The conserved energy and angular momentum as functions of the radius of the orbit of the test particle
are plotted in Figs. \ref{Fig8} and \ref{Fig9} respectively for different values of the parameter $\gamma$. The qualitative behaviour of these functions is also manifestly different in the three regimes of the parameter mentioned above.

\subsubsection{Case $\gamma \in (0,\frac{1}{\sqrt{5}})$}
In this range of parameter $r_+$ and $r_-$ do not take real values and we have
\begin{equation}
r_{sing}>r_{ph} \; ,
\end{equation}
meaning also that the photon sphere does not exist.
This implies that the stable orbits can exist from infinity to the singularity:
\begin{eqnarray}
r \in (r_{sing},\infty) \; .
\end{eqnarray}

\begin{figure}
\includegraphics[width=0.5\textwidth]{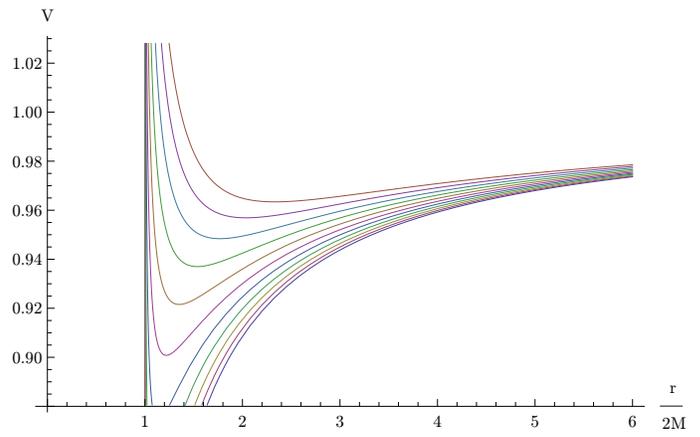}
\caption{\label{Fig10}
The effective potential $V$ is plotted against radial coordinate $r$ in units of $\frac{1}{2M}$ for $\gamma=0.3 \in(0,\frac{1}{\sqrt{5}})$ for different values of $L$. The effective potential admits only a  minimum implying that unstable orbits in this case do not exist.
}
\end{figure}

In Fig.\ref{Fig10} the general behaviour of the effective potential for this range of parameter is shown. The effective potential admits only one minimum. Since the potential does not admit maximum, unstable circular orbits do not exist. This is consistent with the existence of stable circular orbits everywhere as discussed above.

The qualitative behaviour of the functions $E(r)$ and $L(r)$, as shown in Figs. \ref{Fig8} and \ref{Fig9}, is decreasing as the radial coordinate decreases from infinity and vanishes at the singularity.

\subsubsection{Case $\gamma \in (\frac{1}{\sqrt{5}},\frac{1}{2})$}
In this range of parameters we get the following relation
\begin{equation}
r_+ >r_{-}>r_{sing}>r_{ph} \; ,
\end{equation}
which means that no photon sphere is present in the spacetime and there are two distinct regions where circular geodesics can exist, namely
\begin{equation}
r \in (r_{+},\infty) \; \text{and} \;
r \in (r_{sing},r_{-})
\end{equation}
where the first region extends from infinity and the second region extends until to the singularity.

\begin{figure}
\includegraphics[width=0.5\textwidth]{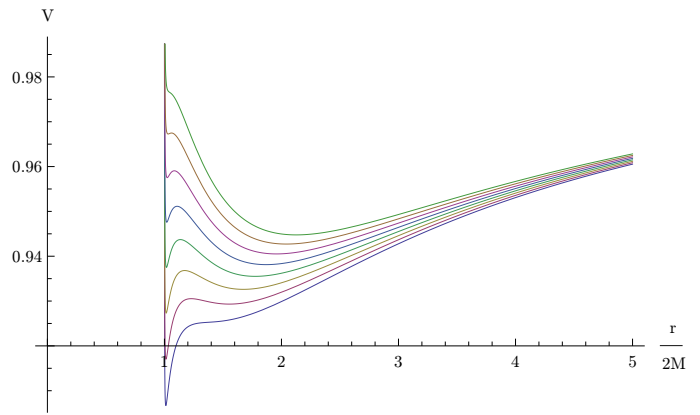}
\caption{\label{Fig11}
The effective potential for radial motion $V$ is plotted as a function of radial coordinate $r$ in units of $\frac{1}{2M}$ for $\gamma=0.47 \in (\frac{1}{\sqrt{5}},\frac{1}{2})$ for different values of $L$. It can be seen that the $V$ admits two minima and one maximum indicating the presence of two regions of stable circular orbits separated by a region of instability.}
\end{figure}

In Fig. \ref{Fig11} we show the general behaviour of the effective potential for this range of parameter values. The effective potential admits two minima. These correspond to the two stable circular orbits in the two regions we discussed above. There is also one local maximum between the two minima which lies in the unstable region between the two stable regions.

A particle reaching the innermost stable circular orbit of the outer disk would plunge in towards the singularity until it reaches $r_-$ where stable orbits are allowed again. It would then shock and circularize its motion spiralling towards the singularity.

The qualitative behaviour of the energy and angular momentum functions $E(r)$ and $L(r)$ is shown in Figs. \ref{Fig8} and \ref{Fig9}.
In this case $E$ and $L$ vanish at the singularity and both have a local maximum and a minimum indicating
again that there are two regions where stable circular geodesics can exist. The two regions would correspond to the places where functions decrease with decreasing $r$, namely from infinity to first minimum and from the maximum to the singularity.
Again the unstable region extends between maximum and minimum, where $E(r)$ and $L(r)$ increase with decreasing $r$.
Since these functions do not blow up at any radius we infer that the photon sphere is absent in this case.

\subsubsection{Case $\gamma \in (\frac{1}{2},\infty)$}
For this range of values of the parameter we have the following relations
\begin{equation}
r_+ >r_{ph}> \text{max}\left\{r_-,r_{sing}\right\} \; ,
\end{equation}
as can be seen from Fig. \ref{Fig7}.
Note that $r_-$ can be either larger or smaller than $r_{sing}$ but in both cases lies below the photon sphere.
The stable circular orbits can exist in the region
\begin{equation}
r \in (r_{+},\infty) \; ,
\end{equation}
which is above the photon sphere. Unstable circular orbits can exist between the photon sphere and $r_+$. No circular orbits can exist below the photon sphere until the singularity.

\begin{figure}
\includegraphics[width=0.5\textwidth]{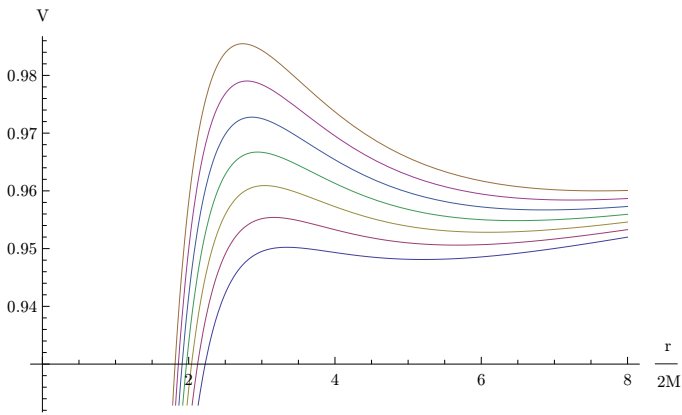}
\caption{ \label{Fig12}
The effective potential for radial motion $V$ is plotted as a function of radial coordinate $r$ (in units of $\frac{1}{2M}$) for $\gamma=1.4 \in (\frac{1}{2},\infty)$ for different values of $L$. It can be seen that the $V$ admits one minimum and one maximum, which corresponds to stable and unstable circular orbit respectively.
}
\end{figure}

In Fig. \ref{Fig12} we show the general behaviour of the effective potential as a function of radius for the parameter in this range. The effective potential admits one maximum and one minimum which correspond to the unstable and stable circular orbits respectively. The unstable circular orbits lie in the region between the photon sphere and $r_{+}$.

The qualitative behaviour of the energy and angular momentum functions $E(r)$ and $L(r)$ shows a minimum and it diverges at a certain radius above the singularity corresponding to the photon sphere (see Figs. \ref{Fig8} and \ref{Fig9}). This is consistent with the above discussion on the structure of geodesics since as $r$ decreases from infinity, the stable circular geodesics can exist up to the minimum, then unstable orbits exist in the region between the minimum and the radial value where $E(r)$ and $L(r)$ blow up, which is a photon sphere. No circular orbits can exist below the photon sphere.

\section{Accretion disk properties}
In the context of astrophysics an accretion disk is a nearly flat disk
of heated gas whose particles slowly spiral onto a central accreting
compact object. Observationally such accretion disks can be detected
by their emitted radiation and the spectrum of this radiation can be a
valuable source of information about the properties of the accretion disk
itself and the central compact object.
In the present framework we approximated the accretion disk with an infinitesimally thin disk of test particles
moving on circular orbits. The gas is then represented by the test particles moving with angular velocity
$\omega=\frac{d\phi}{dt}$ on Keplerian orbits around the central
compact object. Then $E(r)$ and $L(r)$ would
represent the energy and angular momentum of the test particles circling on the $r$-orbit of the disk.
Then the last stable circular orbits represent the limits within which the disk can exist
with the innermost stable circular orbit given by the radius $r_+$ representing the inner edge of the outer disk.
A test particle reaching $r_+$ would then plunge in free fall until it reaches either the singularity or another radius where
stable circular orbits are allowed.
The general relativistic setting for accretion disks around a black hole was developed by Novikov and Thorne
\cite{Novikov} and by Page and Thorne \cite{Page}
for steady state accretion on the equatorial plane onto a rotating black hole and it can be extended to the
scenarios with naked singularities analyzed here.

Page and Thorne assumed the spacetime to be stationary, axially symmetric, asymptotically flat and reflection
symmetric in an equatorial plane. Here both metrics satisfy these requirements once we
consider accretion disks on the equatorial plane (note that the JNW metric is spherically symmetric and thus such a choice
of the plane is arbitrary while the $\gamma$ spacetime is axially symmetric and therefore the equatorial plane is fixed).
Furthermore the choice of the radial coordinate complies with the requirements imposed in \cite{Page} as can be seen
from the asymptotic behaviour of both metrics and from the fact that $r$ reduces exactly to the radial coordinate
of the Schwarzschild metric for $\gamma=1$ in the $\gamma$-metric and to a translation of the Schwarzschild radial coordinate
for $\mu=1$ in the JNW metric.

From the knowledge of the structure equations describing the gas in the disk it is possible to construct the
quantities relevant for observations \cite{Page}.
The radiant energy flux can be written as
\begin{equation}\label{flux}
    f(r)= -\frac{\dot{M}_0}{4\pi\sqrt{-\det{g}}}\frac{\omega_{,r}}{(E-\omega L)^2}\int^r_{r_+}(E-\omega L)L_{,r}dr
\end{equation}
where $\dot{M}_0$ is the mass accretion rate and it is assumed to be constant for steady state accretion.
Note that since both metrics considered here are continuously linked to the Schwarzschild solution
one retrieves precisely the formula for the radiant energy flux in the Schwarzschild spacetime once the
values $\mu=1$ or $\gamma=1$ are chosen.

As an example we studied here the qualitative behavior of the radiant energy flux in the JNW metric and
in the $\gamma$-spacetime in natural units (therefore to obtain the effective flux one would have to
reintroduce $c$ and $G$ in the equation), for different values of the parameters $\mu$ and $\gamma$.

\begin{figure}
\centering
\includegraphics[width=0.5\textwidth]{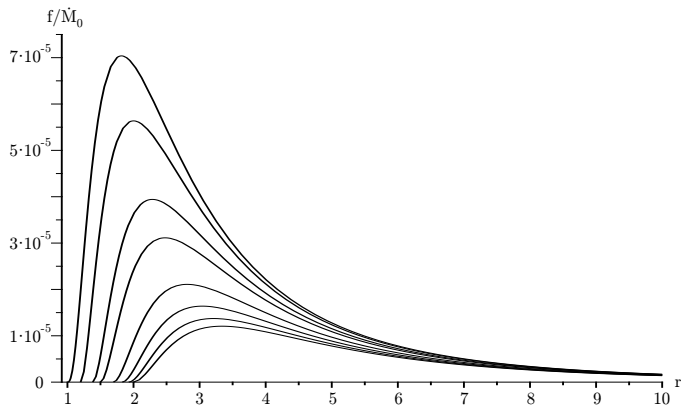}
\caption{\label{Fig13} Radiant energy flux for unit mass accretion rate in natural units for the JNW spacetime with $\mu\leq\sqrt{5}$.
Note that the lowest curve represents the flux for the Schwarzschild black hole.}
\end{figure}

The general feature that emerges for the JNW metric is that when there is an inner boundary for the disk
one obtains a flux qualitatively similar but of greater magnitude with respect to that of a black hole
(see Fig. \ref{Fig13}).

\begin{figure}
\centering
\includegraphics[width=0.5\textwidth]{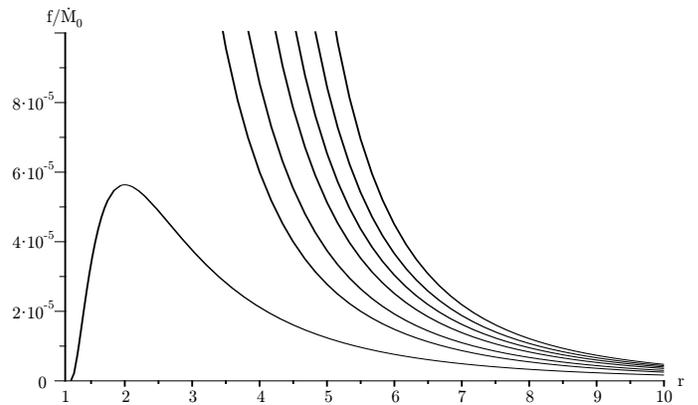}
\caption{\label{Fig14} Radiant energy flux for unit mass accretion
  rate in natural units for
the JNW spacetime with $\mu\geq\sqrt{5}$. Note that the lowest curve
represents the case
$\mu=\sqrt{5}$, with all values greater than $\sqrt{5}$ diverging as
$r$ approaches the singularity.}
\end{figure}

On the other hand when the accretion disk is allowed to extend up to the singularity the
total flux is diverging in the limit of $r$ going to the singularity (see Fig. \ref{Fig14}).
This divergence of flux in the limit of approach to the singularity is clearly unphysical.
Nevertheless we note here that as the outgoing radiation becomes larger, the
outward force acting on the infalling accreting matter will also
become bigger. It can be shown that at a particular radius the
pressure exerted due to the outgoing radiation will balance out the inward gravitational pull of the gas in the disk.
Therefore the accretion disk will not extend below this radius regardless of the presence of a
singularity at the center.
Thus although the circular geodesics as we have shown in this paper
can extend all the way up
to the singularity, it is reasonable to expect that physically the
accretion disk will be terminated
at a finite radius above the singularity.
Nevertheless this is a complex issue that is not entirely well
understood at present
and requires more work and further discussion which is beyond the scope of this paper.

Furthermore since the $\gamma$-metric is not spherically symmetric it should be noted that the axial
symmetry will have some influence on the structure of accretion disks close to the singularity.
In fact it can be shown that close to the center strong forces in the $z$ direction appear in the spacetime.
These forces would disrupt the inflow of particles on the equatorial plane thus deviating them towards the poles,
which could be a mechanism that may in principle describe the formation of high
energy jets from the poles of such compact objects.

\begin{figure}
\centering
\includegraphics[width=0.5\textwidth]{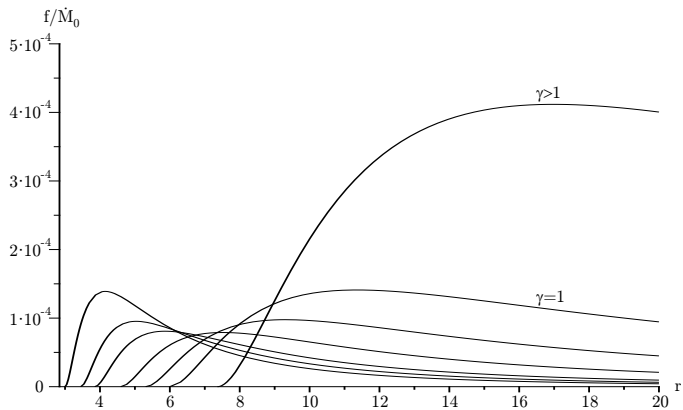}
\caption{\label{Fig15} Radiant energy flux for unit mass accretion
rate in natural
units for the $\gamma$-spacetime with $\gamma\geq 1/2$.}
\end{figure}

\begin{figure}
\centering
\includegraphics[width=0.5\textwidth]{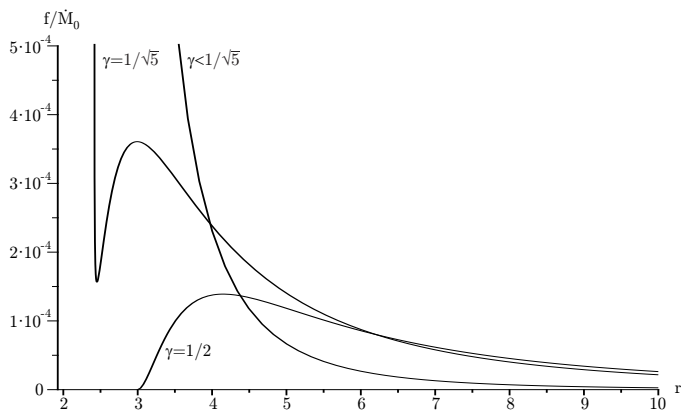}
\caption{\label{Fig16} Radiant energy flux for unit mass accretion
rate in natural units for the $\gamma$-spacetime with $\gamma\leq 1/2$.
Note that the flux diverges for values of $\gamma\leq 1/\sqrt{5}$.}
\end{figure}

In the case of the $\gamma$-metric a similar behaviour is found for
the radiant energy flux
to that of the JNW case. The flux presents a behaviour qualitatively
similar to that
of a black hole when there is an inner edge (see Fig. \ref{Fig15})
and diverges in the cases where the accretion disk extends all the way
to the singularity (see Fig. \ref{Fig16}).

Note that since the total mass as seen by an observer at infinity for
the $\gamma$-metric is $M\gamma$ it can happen for values of
$\gamma<1$
that the flux is smaller than that of a Schwarzschild black hole with
mass $M$. Nevertheless we expect it to be greater than that of a
Schwarzschild
black hole with mass $M\gamma$, which would be the right
spherical source to compare it with.

To summarize we studied in detail the structure of the circular geodesics in
JNW and $\gamma$ spacetimes in the equatorial plane. JNW and
$\gamma$-metric
are obtained by variation of the Schwarzschild black hole by addition of the
massless scalar field and by deformation of spherical symmetry to
oblate/prolate spheroidal geometry respectively. Naively speaking
the event horizon in the Schwarzschild black hole transforms into
a naked singularity. These solutions are described by two parameters,
one related to the mass as in the Schwarzschild case and another
related to the scalar charge ($\mu$) or the deformation ($\gamma$)
and continuously linked to the Schwarzschild spacetime. In fact
the Schwarzschild metric is obtained by setting $\mu=1$ and $\gamma=1$.
The structure of the circular geodesics undergoes a qualitative
change as we vary the values of the extra parameter. In both
spacetimes
the whole range of the parameters can be divided into three regimes.

\begin{itemize}
  \item[-] For $\mu \in(1,2)$ in JNW metric and for $\gamma
    \in(\frac{1}{2},\infty)$
in $\gamma$-metric a photon sphere surrounds the singularity and
stable circular geodesics can exist from a radius above the photon
sphere to infinity. Unstable circular geodesics can exist below
this radius up to the photon sphere. No circular geodesics
can exist below the photon sphere.
  \item[-] For $\mu \in (2,\sqrt{5})$ in JNW metric and for
$\gamma \in (\frac{1}{\sqrt{5}},\frac{1}{2})$ in the $\gamma$-metric,
    stable
circular geodesics can exist from the singularity to a finite radius
and from
a larger radius to infinity. There is a region containing unstable
circular geodesics between the two stable regions. No photon sphere is present.
  \item[-] For $\mu \in(\sqrt{5},\infty)$ in JNW metric and for
$\gamma \in(0, \frac{1}{\sqrt{5}})$ in $\gamma$-metric, the circular
    geodesics
exist everywhere from the singularity to infinity. Again
no photon sphere is present.
\end{itemize}

This study suggests that there is a significant difference as far
as the structure of circular geodesics is concerned in JNW and
$\gamma$ spacetime
containing naked singularity as compared to the Schwarzschild
black hole, which would reflect into their accretion
disk properties.

Similar analysis of circular geodesics in connection with
properties of accretion disks around naked singularities have recently
drawn some attention as we try to understand the nature of compact
objects such as the supermassive ones that exist at the center of galaxies.
Harko and Kovacs
\cite{Harko}
studied a modified Kerr metric with scalar field where, due
to the presence of the scalar field, a naked singularity is present
in the spacetime. They showed that for certain ranges of the
parameters, stable circular orbits can exist up to the singularity,
a result that does not hold for the Kerr black hole
(obtained in the limit of no field).

A similar analysis has been carried out for the Kerr and
Reissner-Nordstrom metrics by Pugliese et al
\cite{accretion2}
where it was shown that for the Kerr naked singularity ($a/M>1$)
unstable circular orbits can extend up to the singularity in some
cases,
though this does not happen for stable circular orbits. Therefore
an accretion disk made of test particles always extends from
a minimum radius to infinity, without any forbidden regions
or regions of instability. The disk can be divided into
substructures based on whether they consist of corotating
and/or counter-rotating particles.
For a naked singularity, depending on the value of $a/M$, the
accretion disk may be divided into: (a) two parts - an inner disk
with co-rotating particles and an outer disk with both co-rotating
and counter-rotating particles, (b) three parts, namely an innermost
disk of counter-rotating particles, a second disk of
co-rotating particles and an external disk with both.
For a Kerr black hole the allowed region of circular orbits,
stable or unstable, terminates outside the horizon. However the
accretion
disk is similar, extending from a minimum radius to infinity
and it can have only two such substructures, a co-rotating inner
disk and an external disk with both kinds of particles; here
the size of the inner disk has an upper limit and can even
go to zero. In the case (a) for naked singularity, the inner
disk has a minimum size but there is no upper limit thus
showing that there is some difference in the accretion disk
structure between Kerr black hole and Kerr naked singularity.
In the Reissner-Nordstrom spacetime a naked singularity is
obtained for $Q/M>1$, where $Q$ is the charge parameter. In this
case there exists a forbidden zone outside the horizon or the
singularity where no circular orbits are allowed. For the black hole
case, the stability region and hence the accretion disk is
continuous from a minimum radius to infinity. In the case of a
naked singularity, there can be two scenarios depending on
the value of $Q/M$, namely (a) an accretion disk extending
from a minimum radius to infinity, or (b) an accretion disk
comprising an inner disk and an outer disk separated by a
zone of instability. Therefore the accretion disk structure
around a Reissner-Nordstrom naked singularity shows some difference
with respect to the black hole case only for a certain
range of the parameter.

\section{Concluding remarks}

The study of circular geodesics provides only the first
tool to model accretion disks around a compact source. In order
to obtain measurable quantities that could eventually be
checked against observations, a detailed analysis of the efficiency
of conversion of mass of the infalling particles into emitted
radiation and the luminosity spectrum of radiation of the disk
are needed.
For example assuming that the accretion disk is in thermodynamical
equilibrium one can approximate the radiation emitted by the disk with a
black body spectrum where the temperature $T$ of the black body is related
to the energy flux by $f(r)=\sigma T^4(r)$ (with $\sigma$ being the Stefan-Boltzmann constant).
Then, considering redshift and inclination of the disk with respect to the observer, it is possible
to evaluate the observed luminosity spectrum of the black body, which is an observable quantity
\cite{Harko}.

It is reasonable to suppose that the pressure of
the outgoing radiation near the singularity will be very strong,
thus balancing the inflow of particles from the accretion disk.
This effect could possibly drive the particles away from the
equatorial plane and it should also be analyzed in some detail.
Furthermore electromagnetic fields and possible perpendicular
forces that could deviate the particles towards the poles
thus creating high energy jets could also be considered.

Nonetheless the analysis such as given here makes it
clear that if naked singularities
which are hypothetical astrophysical objects do exist in the
universe, they would bear an observational signature considerably
different from that of a black hole of the same mass.
Therefore it is quite possible that in the foreseeable future
we will be able to devise some observational tests to check
if such exotic objects are indeed present somewhere
in the universe.

Acknowledgement: ANC thanks TIFR for hospitality during
VSRP2011 program during which this work was done.

\end{document}